\documentclass[reprint,amsmath,amssymb,aps,prb,longbibliography]{revtex4-2}

\usepackage{natbib}
\usepackage{graphicx,bm,physics,hyperref}
\hypersetup{breaklinks=true,colorlinks = true,allcolors = blue }
\usepackage{mathtools}

\usepackage{amsthm}
\newtheorem*{theorem*}{Theorem}

\usepackage{environ} 
\NewEnviron{eqn}{
\begin{align}
    \begin{split}
        \BODY
    \end{split}
\end{align}
}

\newcommand{\bb}{\mathcal{B}}
\newcommand{\uu}{\mathcal{U}}
\newcommand{\cc}{\mathcal{C}}
\newcommand{\vv}[1]{\boldsymbol{#1}} 
\newcommand{\p}[2]{P_{#1}^{#2}}
 
\DeclarePairedDelimiter\floor{\lfloor}{\rfloor}

\begin{document}
\title{Topological Triviality of Flat Hamiltonians}

\author{Pratik Sathe}
\email{psathe@ucla.edu}
\author{Rahul Roy}
\email{rroy@physics.ucla.edu}
\affiliation{Mani L. Bhaumik Institute for Theoretical Physics, Department of Physics and Astronomy, University of California at Los Angeles, Los Angeles, CA 90095}

\date{\today}

\begin{abstract}
Landau levels play a key role in theoretical models of the quantum Hall effect.
Each Landau level is degenerate, flat and topologically non-trivial.
Motivated by Landau levels, we study tight-binding Hamiltonians whose energy levels are all flat.
We demonstrate that in two dimensions, for such Hamiltonians, the flat bands must be topologically trivial.
To that end, we show that the projector onto each flat band is necessarily strictly local.
Our conclusions do not need the assumption of lattice translational invariance. 
\end{abstract}

\maketitle
\setcounter{footnote}{0} 

\emph{Introduction.}--
Topological phases of matter are a cornerstone of modern condensed matter physics.
Beginning with the discovery of the quantum Hall effects (QHE)~\cite{KlitzingNew1980, TsuiTwo1982}, topology has played an important role in understanding and predicting novel phases of matter ranging from topological insulators and superconductors to topologically ordered phases of matter~\cite{HasanColloquium2010a,WenColloquium2017}.

The Hall conductivity of a filled Landau level or a Bloch band is proportional to a topological invariant called the Chern number~\cite{ThoulessQuantized1982a,NiuQuantized1985a,AvronQuantization1985}.
Much like Landau levels, a Bloch band that is flat or dispersionless and which has a non-zero Chern number can host fractional quantum Hall phases when subject to interactions.
Such systems are known as fractional Chern insulators (FCIs) and hold significant experimental appeal since they have the potential to exhibit fractional QHE at small (or even zero) magnetic fields and high temperatures~\cite{XieFractional2021}.
Consequently, a substantial amount of research has focused on the study of topological flat-band models (see Refs.~\cite{ParameswaranFractional2013,BergholtzTopological2013a,LiuRecent2023a} for reviews).

Hamiltonians with flat bands are also interesting because of numerous other phenomena they exhibit~\cite{LeykamArtificial2018}.
The effects of interactions are pronounced in flat-band Hamiltonians, and hence they serve as a platform to explore novel correlated phases of matter.
In addition to FCIs, some prominent examples include unconventional superconductivity in bilayer graphene~\cite{CaoUnconventional2018,VolovikGraphite2018,LuSuperconductors2019, YankowitzTuning2019}, flat-band superfluidity~\cite{PeottaSuperfluidity2015,XieTopologyBounded2020} and flat-band ferromagnetism~\cite{TasakiNagaoka1998}.

Various no-go theorems restrict the possibility of topological flat bands.
Strict localization of a band projector, a condition more restrictive than band flatness, was shown to imply a vanishing Hall conductance~\cite{BezrukavnikovLocalization2019}.
Further, if a strictly local (SL) Hamiltonian has a flat band, then it must have a Chern number of zero~\cite{Chenimpossibility2014, ReadCompactly2017}.
These results apply to Hamiltonians with lattice translational invariance (LTI).
It has also been shown that local commuting projector Hamiltonians cannot exhibit the QHE~\cite{KapustinLocal2020}.
The non-interacting limit of such a model has a flat band which is spanned by an orthogonal basis of compactly supported wavefunctions~\cite{SatheCompactly2021, SatheCompact2023}.

Non-interacting Hamiltonians that exclusively have flat bands, which we call flat Hamiltonians, have also attracted significant attention.
Flat Hamiltonians can exhibit unconventional behavior when interactions are introduced.
Some examples include many-body localization in translationally-invariant Hamiltonians~\cite{DanieliManybody2020,KunoFlatband2020}, non-linear caging~\cite{DanieliNonlinear2021,GligoricNonlinear2019} and transport via two-particle bound states~\cite{TovmasyanPreformed2018,PelegriInteractioninduced2020}.
Furthermore, the spectrum of a flat Hamiltonian is similar to Landau levels since all the energies are flat and degenerate.
Due to this resemblance, topological flat Hamiltonians that have SL hoppings seem to be particularly promising as platforms for fractional QHE.
Indeed, many of the initial proposals for FCIs were Hamiltonians with all bands being approximately flat~\cite{NeupertFractional2011a,SunNearly2011,TangHighTemperature2011a,WangNearly2011}.

In this Letter, we prove some unique properties of SL flat Hamiltonians with exactly flat bands and establish a no-go theorem. 
First, we show that the projection operators associated with all the flat bands in a SL flat Hamiltonian are also SL, regardless of lattice dimension or the system having LTI.
Next, we show that the spectrum of such a Hamiltonian is unchanged even with twisted boundary conditions.
By utilizing properties unique to SL projectors, we show that each flat band in a flat Hamiltonian has a Chern number of zero.
(We compute the many-body Chern number defined in terms of the twist angles~\cite{NiuQuantized1985a}.)
More generally, our statement applies to any Hamiltonian that has a highly degenerate spectrum, regardless of whether it has LTI.
The no-go theorem is summarized below.
\begin{theorem*}
Consider a SL tight-binding Hamiltonian defined on a 2d system of size $L_{x}\times L_y$ cells with periodic boundary conditions.
Let $n$ denote the number of distinct energies and $R$ denote the maximum hopping range of the Hamiltonian.
If $L_{x,y} \geq 3nR$, then the Chern number associated with each energy is $0$.
\end{theorem*}
We note that within the context of flat Hamiltonians with LTI, the inequality above is broadly satisfied.  
However, exotic models (such as those in Ref.~\cite{ScaffidiExact2014a} which have an extensive number of flat bands) that do not have this property can evidently be topological.

\emph{Notation and Setup.}---
We consider 2d tight-binding Hamiltonians with orbitals $\ket{\vv r, \alpha}$ where $\vv r = (x,y)$ denotes cell position and $\alpha=1,\dotsc,N$ denotes different orbitals in a unit cell.
We will restrict our discussion to systems that lie on a torus (i.e. those that satisfy periodic boundary conditions).
Without loss of generality, we consider square lattices of size $L_x \times L_y$ with lattice constant $1$. 

The concepts of strict localization of operators and wavefunctions will play an important role in our arguments. 
A wavefunction is compactly supported if it has non-zero support only on a finite region of the lattice.
On the other hand, an operator $\mathcal{O}$ in an infinite sized system is said to be strictly local (SL) if $\bra{\vv r, \alpha} \mathcal{O} \ket{\vv r', \beta} = 0$ (for any $\alpha, \beta$) whenever $|\vv r - \vv r'| > R$ for some finite number $R$. 
We call the smallest value of $R$ for which this is true the maximum hopping range of the operator.
For finite sized systems, this condition is satisfied by every operator.
Therefore, we instead propose the following definition for finite-sized systems on a torus: an operator is called SL if it does not connect any two cells that are maximally separated from each other.
This is equivalent to the conditions $2R+2 \leq L_{x/y}$.

The central focus of this manuscript is on flat Hamiltonians, which we now define.
We recall that in an infinite sized system, the spectrum of a Hamiltonian with LTI consists of a finite number of continuous bands following Bloch's theorem.
The energy eigenvalues of a Hamiltonian on a finite system are necessarily discrete but are still arranged in the form of bands if the Hamiltonian is LTI. 
In particular, each band is spanned by Bloch wavefunctions at a set of discrete values of the crystal momentum that are equal in number to the system size.
If a Hamiltonian with LTI only has flat bands, then we call it a flat Hamiltonian.
Thus, each energy level has a degeneracy equal to the system size.
In the case of Hamiltonians without LTI, we consider Hamiltonians that have a similar property of possessing highly degenerate energy levels.
In a slight abuse of notation, we will refer to the distinct energy levels of such Hamiltonians as flat bands even in the absence of LTI.
Regardless of LTI, all our results are derived for Hamiltonians and system sizes which satisfy $3nR \leq L_{x,y}$, where $n$ denotes the number of distinct energies or flat bands, and $R$ denotes the maximum hopping range of $H$~\footnote{Strictly speaking, in addition to the condition $L_{x,y} \geq 3nR$, we also require that the Hamiltonian satisfy $L_{x,y} \geq 2nR +2$ to ensure that all the projectors are SL. Except for the special (and uninteresting) case of $nR\in \{0,1\}$, it is always true that $3nR \geq 2nR +2$, so that we simply require $L_{x,y} \geq 3nR$.}.

Each energy eigenspace of a Hamiltonian may be associated with an orthogonal projector, i.e., a Hermitian operator $P$ which satisfies $P^2=P$. The locality properties of  projectors associated with energy eigenspaces will form an important part of our subsequent discussion.  

We will also find it useful to use a simple kind of dimensional reduction. 
Specifically, any 2d SL operator can always be regarded as a 1d SL operator by grouping together all the lattice sites in every row (or column) into a single 1d lattice site. 
In particular, a 2d next-nearest-neighbor (NNN)-hopping projector $P$ (i.e. one with $R=\sqrt 2$) is also a nearest-neighbor (NN)-hopping projector (with $R=1$) on a 1d lattice with one truncated dimension.

We will find it useful to define `hopping matrices' for 1d SL projectors.
Specifically, we define $\p i x$ for a SL projector $P$ as follows:
\begin{align}
    \p i x \coloneqq \bra{x+i} P \ket{x}. \label{eq:defn_hopping_matrices}
\end{align}
The number of rows or columns in this matrix is equal to the number of orbitals in each cell.
We note that such hopping matrices can also be defined for 2d SL projectors by regarding them as 1d projectors using the dimensional reduction described above.

\emph{Flat Hamiltonians have strictly local projectors.}---
We will now show that if a SL Hamiltonian is flat, then all the projectors are also SL~\footnote{In Ref.~\cite{DanieliNonlinear2021}, it was stated without proof that 1d SL flat Hamiltonians with LTI must have SL band projectors. [See the discussion below equation (A22) in that paper.] While this statement is true, we believe that it is not trivial to prove. Hence by proving this statement, we fill a gap in their proof for the statement that each band in a 1d flat Hamiltonian can be spanned by compact Wannier functions. Furthermore, our proof regarding the projectors is not restricted to 1d systems, or to systems with LTI.}. 
Furthermore, if the maximum hopping range of the Hamiltonian is $R$ and the number of distinct energies is $n$, then the maximum hopping range of each projector is less than or equal to $nR$.

To that end, consider a Hamiltonian $H$ with energy eigenvalues $E_1 < E_2 < \dotsc < E_n$.
Let the corresponding projectors be denoted by $P_1,P_2,\dotsc, P_n$. 
Without loss of generality, all the energies can be chosen to be non-zero~\footnote{Adding a constant to the Hamiltonian has no effect on the projectors.}. 
Taking the $m^{th}$ power of $H$, we get
\begin{align}
    H^m &= \sum_{i=1}^n E_i^m P_i,\label{eq:powers_of_H}
\end{align}
for all integers $m\geq 1$. This follows from the fact that $P_i P_j = P_j P_i = P_i \delta_{i,j}$ for all $i,j$.

For any $m$, $H^m$ has a maximum hopping range of no more than $mR$. Therefore, each of $H, \dotsc, H^n$ has vanishing matrix elements between any two locations separated by a distance greater than $nR$. 
For concreteness, consider orbitals $\alpha$ and $\beta$ at any two cells $\vec{r}_1$ and $\vec{r_2}$ respectively, such that $|\vec r_1-\vec r_2|>nR$.
Let  $j\equiv (\vec r _1, \alpha)$ and $k \equiv (\vec r _2, \beta)$ label the orbitals.
Equating the $(j,k)$ matrix elements of both sides of equations \eqref{eq:powers_of_H} for $m =1,\dotsc, n$ gives us
\begin{align}
    \underbrace{\begin{pmatrix}
        E_1 & E_2 &\dotsc &E_n \\
        E_1^2 & E_2^2 &\dotsc & E_n^2 \\
        \vdots & & & \vdots \\
        E_1^n & E_2^n &\dotsc &E_n^n 
    \end{pmatrix}	}_{M}
    \begin{pmatrix}
        (P_1)_{j,k}\\
        (P_2)_{j,k}\\
        \vdots\\
        (P_n)_{j,k}
    \end{pmatrix} = 
    \begin{pmatrix}
        0\\
        0\\
        \vdots\\
        0
    \end{pmatrix} \label{eq:homogeneous_linear_equations}
\end{align}
The only solution of \eqref{eq:homogeneous_linear_equations} is the zero vector since
\begin{align*}
    \det M&= \prod_{i=1}^n E_i   \prod_{1\leq j < k \leq n} (E_j - E_k) \neq 0,
\end{align*}
where the second product series is the Vandermonde determinant.
Let $p_i$ denote the maximum hopping range of projector $P_i$.
We have thus shown that $p_i \leq nR$ for each $i$.
We note that all the conclusions so far are valid independent of the system dimensionality and regardless of LTI, even though we will focus on 2d systems below.

For systems of infinite size and finite $n$, all the projectors are thus SL.
However, for finite-sized systems, the projectors may not be SL (in the sense defined above for finite-sized systems) if $nR$ is comparable to the system size.
We note that $q\coloneqq \max \{R, p_1,\dotsc, p_n\}$ is such that $q\leq nR$.
All the projectors are SL if $2q+2 \leq L_{x/y}$.
This is therefore guaranteed if $2nR+2\leq L_{x/y}$.

In the subsequent sections, the following observation will be useful.
If we group the unit cells so that $q \times q$ primitive cells form a supercell, then all the projectors and the Hamiltonian are NNN-hopping operators~\footnote{If $L_x$ is not divisible by $q$, then we choose the first $\floor*{L_x/q}-1$ number of supercells along $\hat x$ to consist of $q$ cells each (where $\floor*{.}$ denotes the floor function), while the last supercell will comprise $L_x - q (\floor*{L_x/q}-1)$ number of primitive cells. We implement a similar grouping along $\hat y$ if $L_y$ is not divisible by $q$.}.

\begin{figure}[t]
    \centering
    \includegraphics[width=0.75\columnwidth]{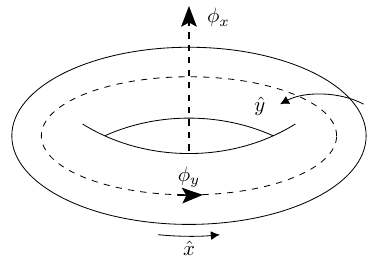}
    \caption{Periodic boundary conditions are imposed on the system, so that it lies on a torus.
    Two magnetic fluxes $\phi_{x}$ and $\phi_y$ are inserted through the two holes of the torus. 
    The directions $\hat x$ and $\hat y$ are shown for reference.}
    \label{fig:fluxes_schematic}
\end{figure}

\emph{Spectrum is unchanged upon flux insertion.}---
The spectrum of a non-flat Hamiltonian generally changes when magnetic fluxes are threaded through the two holes of a torus as shown in Fig.~\ref{fig:fluxes_schematic}. 
However, for flat Hamiltonians which have SL projectors (and which further satisfy $L_{x,y} \geq 3q$ and $L_{x,y}\geq 2q+2$), the spectrum is unchanged even after threading any amount of flux, as we will show below.

Let us first consider the case when only $\phi_x$ is inserted, so that $H$ changes to $H(\phi_x)$ following Peierls substitution~\cite{PeierlsZur1933}.
We choose a gauge for the substitution that corresponds to twisted boundary conditions (TBCs)~\cite{NiuQuantized1985a}, so that only those hopping elements that cross the boundary at $x=L_x$ are modified.
In other words, every hopping elements that connects a cell with an $x$-coordinate in $\{L_x-R+1, \dotsc, L_x\}$ to a cell with an $x$-coordinate in $\{1,2,\dotsc,R\}$ is multiplied by a phase of $e^{-2\pi i \phi_x/\Phi_0}$, while hoppings in the reverse direction get multiplied by $e^{2\pi i \phi_x/\Phi_0}$.
($\Phi_0=h/e$ is the magnetic flux quantum.)
The rest of the hopping elements are unchanged in the substitution.

Similarly, we can implement TBCs for all the $P_i$ as well, if they are SL.
Denoting the post-transformation operators by $P_i(\phi_x)$s, we note that
\begin{align}
    H(\phi_x) = \sum_{i=1}^N E_i P_{i}(\phi_x). \label{eq:H_after_phi_x}
\end{align}
While it is not immediately obvious, this equation is actually a spectral decomposition of $H(\phi_x)$ when $L_{x} \geq 3q$, as shown in Appendix~\ref{appendix:unchanging_spectrum_proof}.

This fact can also be used to obtain a spectral decomposition for the Hamiltonian $H(\phi_x,\phi_y)$, the Hamiltonian with both the fluxes are inserted, if $L_y \geq 3q$ and $L_y\geq 2q+2$.
The key idea is to treat $H(\phi_x)$ as the `initial Hamiltonian' and repeat the steps above. 
Specifically, we obtain $H(\phi_x,\phi_y)$ by applying TBCs to $H(\phi_x)$ along $\hat y$.
Transforming the projectors similarly, so that $P_i(\phi_x) \rightarrow P_i(\phi_x,\phi_y)$, we get
\begin{align}
    H(\phi_x,\phi_y) = \sum_{i=1}^N E_i P_{i}(\phi_x,\phi_y) \label{eq:H_after_phi_x_and_phi_y}.
\end{align}
Applying the proof from Appendix~\ref{appendix:unchanging_spectrum_proof} again, we conclude that \eqref{eq:H_after_phi_x_and_phi_y} is also a spectral decomposition.
Thus, flat Hamiltonians have the interesting property of having an unchanging spectrum if $L_{x,y} \geq 2q+2$ and $L_{x,y} \geq 3q$.
Both these conditions are satisfied if $L_{x,y} \geq 3nR$ and $L_{x,y}\geq 2nR+2$.

\emph{Many-Body Chern Number.}---
We now turn to the computation of the Chern number.
Let us recall that in the absence of interactions, the transverse conductivity of a system with a Hamiltonian that has LTI is proportional to the TKNN invariant~\cite{ThoulessQuantized1982a}, i.e. the Chern number of the occupied Bloch bands.
More generally, the transverse conductivity is proportional to the Chern number of the (gapped) many-body ground state as a function of twist angles or equivalently, the fluxes $\phi_{x,y}$~\cite{NiuQuantized1985a}. 
This many-body Chern number is expressed as
\begin{align}
    C &= \frac{1}{2\pi} \int_0^{\Phi_0}  \int_0^{\Phi_0} \dd \phi_x \ \dd \phi_y \curl i \bra \Psi \ket {\nabla_{\phi} \Psi}, \label{eq:NTW_formula}
\end{align}
where $\ket{\Psi(\phi_x,\phi_y)}$ is the many-body ground state of $H(\phi_x,\phi_y)$.
Even for finite systems, the transverse conductivity has been shown to be proportional to this Chern number up to exponentially small errors in the system size~\cite{HastingsQuantization2015, KudoManyBody2019}.
We can straightforwardly generalize \eqref{eq:NTW_formula} in order to define a Chern number for an isolated flat band or energy in a flat Hamiltonian.
Specifically, we use expression Eq.~\eqref{eq:NTW_formula} but with $\ket{\Psi}$ taken to be the `flat band wavefunction', i.e. the many-body wavefunction corresponding to only the flat band states being fully occupied, with the rest of the energies being unoccupied.
We will henceforth use this definition.

\emph{Many-body wavefunctions as a function of the fluxes.}--- 
Let us proceed with the task of computing the Chern number corresponding to a flat band in a flat Hamiltonian.
Let the energy of the flat band be $E$ and the corresponding projector, $P$.
We assume that $L_{x,y}\geq 3nR$ so that the spectrum of a flat Hamiltonian is unaffected by the fluxes as shown above.
Thus, for any $\phi_x,\phi_y$, while the state of the system corresponding to that energy level being fully filled is unique, the corresponding wavefunction $\ket{\Psi(\phi_x,\phi_y)}$ has a phase ambiguity.
We will now provide a prescription that fixes the phase to obtain a global expression for $\ket{\Psi(\phi_x,\phi_y)}$ that can then be used to compute the Chern number.

We will start with a brief outline of our approach.
First, we obtain a particular type of orthogonal basis of single-particle wavefunctions spanning the energy $E$ subspace when no fluxes are inserted.
The Slater determinant then yields $\ket{\Psi}$ at zero flux.
Next, we consider the case when only $\phi_x$ is inserted, so that $P$ transforms to $P(\phi_x)$.
We show that wavefunctions obtained for $\phi_x=0$, when modified appropriately, are eigenstates of $P(\phi_x)$ and thus span $E$ at flux $\phi_x$.
The corresponding Slater determinant yields $\ket{\Psi(\phi_x)}$, which by construction is shown to be a smooth function of $\phi_x$.
In the last step, we insert flux $\phi_y$ in addition to $\phi_x$.
Specifically, for each value of $\phi_x$, we follow a similar procedure to obtain a smooth family of wavefunctions as a function of $\phi_y$.
This finally results in a global expression for $\ket{\Psi(\phi_x,\phi_y)}$.

For the remainder of the manuscript we will work within the size $q\times q$ supercell representation described previously.
Thus, $H$ and all $P_i$s are then NNN operators in this representation, and we have a supercell grid of a size of at least $3\times 3$ since we assumed $L_{x,y}\geq 3nR$.
The wavefunctions obtained in this representation can always be expressed in the original primitive cell representation at the end.

The first step involves obtaining the flat band many-body wavefunction before inserting $\phi_x$. 
First, we demote the $y$ index to an orbital index so that $P$ can be regarded as a NN projector on a 1d lattice with position coordinate $x$.
Next, we will use the property that the image of a 1d SL projector can always be spanned by an orthogonal basis of compactly-supported wavefunctions~\cite{SatheCompactly2021,SatheCompact2023}.
Specifically, we use the construction of generalized Wannier functions presented in Ref.~\cite{SatheCompact2023} for 1d NN projectors.
The wavefunction so obtained are of two types: `monomers' and `dimers'.
A monomer has support on only one cell, while a dimer has support only on two consecutive cells.
(In the original 2d picture, each such wavefunction is strictly localized along $\hat x$ but is possible delocalized along $\hat y$.)

Monomer-dimer bases have an important property which we call the `support property'.
Specifically, whenever two wavefunctions from such a basis have non-zero support at a common cell, say $x_0$, then their supports at $x_0$ are also mutually orthogonal.

We note further that a monomer-dimer basis so obtained is not unique.
For instance, multiplying any monomer or dimer in a given basis by a phase itself results in a different basis.
Less trivial possibilities arise whenver any $\p 1 x$ hopping matrix [defined in \eqref{eq:defn_hopping_matrices}] has degenerate singular values; see Ref.~\cite{SatheCompact2023}.
However, for our purpose, it suffices to choose any valid monomer-dimer basis for $P$.
For convenience, let us denote the chosen basis by a set $\bb$.
We obtain the flat band many-body wavefunction $\ket{\Psi}$ at zero flux by computing the Slater determinant of $\bb$.

\begin{figure}[t]
    \centering
    \includegraphics[width=0.75\columnwidth]{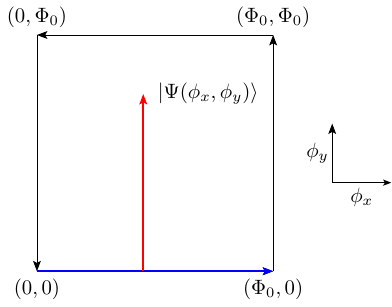}
    \caption{The $(\phi_x,\phi_y)$ parameter space with $\phi_x,\phi_y \in [0,\Phi_0]$. For any pair of values $(\phi_x,\phi_y)$, wavefunction $\ket{\Psi(\phi_x,\phi_y)}$ is obtained in two steps.
    First, we obtain a family of wavefunctions along the blue horizontal line.
    Next, for any value of $\phi_x$, we obtain a family of wavefunction as a function of $\phi_y$.}
    \label{fig:Psi_torus}
\end{figure}

Next, we insert flux $\phi_x$ so that $P$ now transforms to $P(\phi_x)$.
The hopping matrices of $P$ that connect $x=0$ and $x=1$ get transformed, with the rest being unaffected.
The substitution can be summarized as:
\begin{eqn}
        \p 1 0 &\xrightarrow{\phi_x} e^{-2\pi i \phi_x/\Phi_0}\p 1 0 \\
        \text{and }\p {-1} 1 &\xrightarrow{\phi_x}  e^{2\pi i \phi_x/\Phi_0}\p {-1} 1. \label{eq:projector_change_after_phi}
\end{eqn}
While $\bb$ is not an orthonormal basis for the image of $P(\phi_x)$, a simple modification of it does span the image.
Consider two subsets of $\bb$, namely $\cc$ which consists only of those dimers that have support at both $x=0$ and $x=1$, and $\uu$ that consists of all other wavefunctions from $\bb$, so that $\bb=\uu \cup \cc$.
It is easy to see that every element of $\uu$ is also an eigenvector of $P(\phi_x)$ (with an eigenvalue of $1$), since $P(\phi_x)$ is the same as $P$ except for hopping elements between $x=0$ and $x=1$.
However, the elements of $\cc$ are not eigenvectors of $P(\phi_x)$. 
They can however be modified as follows.
Consider a wavefunction $\ket{w} \in \cc$.
Let
\begin{align}
    \ket{w}=\alpha \ket{x=0,\psi_0} + \beta \ket{x=1,\psi_1},
\end{align}
for some support vectors $\ket{\psi_{0,1}}$. 
We now define $\ket{w_{\phi_x}}$ as follows:
\begin{align}
    \ket{w_{\phi_x}} \coloneqq \alpha \ket{x=0,\psi_0} + e^{-i2\pi \phi_x/\Phi_0} \beta \ket{x=1,\psi_1}.
\end{align}
Using \eqref{eq:projector_change_after_phi} and the fact that $P\ket{w}=\ket{w}$, we conclude that $P(\phi_x)\ket{w_{\phi_x}}=\ket{w_{\phi_x}}$.
Let us denote by $\cc_{\phi_x}$ the set obtained after modifying every element of $\cc$ in this way, and define $\bb_{\phi_x} \coloneqq \uu \cup \cc_{\phi_x}$.
Then, using the support property of monomer-dimer bases, we furthermore conclude that $\mathcal{B}_{\phi_x}$ is an orthogonal basis (and actually a monomer-dimer basis) for the image of $P(\phi_x)$.
For any value of $\phi_x$, the Slater determinant of $\bb_{\phi_x}$ thus yields $\ket{\Psi(\phi_x)}$.
Clearly, $\ket{\Psi(\phi_x)}$ is a smooth function of $\phi_x$.
Furthermore, since $\bb_{\phi_x=\Phi_0}$ is the same as $\bb_{\phi_x=0}$, we also find that $\ket{\Psi(\phi_x=0)} = \ket{\Psi(\phi_x=\Phi_0)}$.

Having obtained $\ket{\Psi(\phi_x)}$, we will now discuss the last step, i.e. obtaining the many-body wavefunction when $\phi_y$ is also inserted.
To that end, for any value of $\phi_x$, we first set $H(\phi_x)$ to be the `initial' flat Hamiltonian and then apply the entire procedure discussed above, but for the flux $\phi_y$.
[We thus obtain monomer-dimer states for each band projector $P_i(\phi_x,\phi_y)$, with the states being strictly localized along $\hat y$.]
The Slater determinant of these wavefunctions, denoted $|\tilde \Psi_{\phi_x} (\phi_y) \rangle$, corresponds to the flat band at those flux values.
Since the many-body state corresponding to the flat band is unique for any flux values, $\ket{\Psi(\phi_x)}$ and $|\tilde \Psi_{\phi_x} (\phi_y=0) \rangle$ must be the same up to a phase, so that $\ket{\Psi(\phi_x)} = e^{i\theta_{\phi_x}} |\tilde \Psi_{\phi_x} (\phi_y=0) \rangle$ for some $\theta_{\phi_x} \in \mathbb{R}$.
Finally we define $\ket{\Psi(\phi_x, \phi_y)} \coloneqq e^{i\theta_{\phi_x}} |\tilde \Psi_{\phi_x} (\phi_y) \rangle$.

We find that by construction, for all values of $\phi_x$ and $\phi_y$,
\begin{eqn}
    \ket{\Psi(\phi_x, 0)} &= \ket{\Psi(\phi_x, \Phi_0)}, \\
    \text{and }\ket{\Psi(0, \phi_y)} &= \ket{\Psi(\Phi_0, \phi_y)}. \label{eq:Psi_prop_bndy}
\end{eqn}

Using Stokes' theorem, we can write \eqref{eq:NTW_formula} as
\begin{align}
    C &= \frac{1}{2\pi} \oint _\Gamma  i\bra{\Psi}\ket{\nabla_{\vv \phi} \Psi} . \dd \vv \phi, \label{eq:line_integral}
\end{align}
where $\Gamma$ is the boundary $(0,0) \rightarrow (\Phi_0,0) \rightarrow (\Phi_0,\Phi_0) \rightarrow (0,\Phi_0) \rightarrow (0,0)$.

From \eqref{eq:Psi_prop_bndy}, we conclude that the line integral contributions from $(0,0) \rightarrow (\Phi_0,0)$ and $(\Phi_0,\Phi_0) \rightarrow (0,\Phi_0)$ cancel each other, and so do those from $(\Phi_0,0) \rightarrow (\Phi_0,\Phi_0)$ and $(0,\Phi_0) \rightarrow (0,0)$.
Consequently, $C=0$~\footnote{Let us note that while our proof of Chern triviality was stated for flat bands in flat Hamiltonians, all the steps can also be used to prove that the Chern number associated with an isolated degenerate energy (or a flat band) that has SL projector is zero, regardless of whether the rest of the bands are flat.}.

\emph{Conclusions.}---
We showed that all bands of a two dimensional strictly local (SL) flat Hamiltonians have a Chern number of zero.
To that end, we showed that each band in such a system is described by a SL projector.
We also showed that the spectrum of such Hamiltonians is unchanged even after flux threading in a toroidal geometry.
We demonstrated all our results without the requirement of lattice-translational invariance (LTI) thereby going beyond existing no-go theorems concerning flat bands.
Furthermore, we clarified the role of a subtle condition that allows for a topological flat Hamiltonian in finite systems. 

An important step of our proof involved showing that each band projector is strictly local.
Since projection operators are equal-time Green's functions in non-interacting systems, generalizations to interacting cases might be possible.
Our no-go theorem was proven for system sizes that are greater than a lower bound expressible in terms of the number of energies and maximum hopping range of the Hamiltonian.
It would be interesting to obtain an improved lower bound.

\emph{Acknowledgements.}---
We thank Adrian Culver for providing helpful comments on the manuscript.
P.S. and R.R. acknowledge financial support from the University of California Laboratory Fees Research Program funded by the UC Office of the President (UCOP), grant number LFR-20-653926. 
P.S. acknowledges financial support from the Center for Quantum Science and Engineering Fellowship (UCLA) and the Bhaumik Graduate Fellowship (UCLA).

\bibliography{paper}

\begin{thebibliography}{44}%
\makeatletter
\providecommand \@ifxundefined [1]{%
 \@ifx{#1\undefined}
}%
\providecommand \@ifnum [1]{%
 \ifnum #1\expandafter \@firstoftwo
 \else \expandafter \@secondoftwo
 \fi
}%
\providecommand \@ifx [1]{%
 \ifx #1\expandafter \@firstoftwo
 \else \expandafter \@secondoftwo
 \fi
}%
\providecommand \natexlab [1]{#1}%
\providecommand \enquote  [1]{``#1''}%
\providecommand \bibnamefont  [1]{#1}%
\providecommand \bibfnamefont [1]{#1}%
\providecommand \citenamefont [1]{#1}%
\providecommand \href@noop [0]{\@secondoftwo}%
\providecommand \href [0]{\begingroup \@sanitize@url \@href}%
\providecommand \@href[1]{\@@startlink{#1}\@@href}%
\providecommand \@@href[1]{\endgroup#1\@@endlink}%
\providecommand \@sanitize@url [0]{\catcode `\\12\catcode `\$12\catcode
  `\&12\catcode `\#12\catcode `\^12\catcode `\_12\catcode `\%12\relax}%
\providecommand \@@startlink[1]{}%
\providecommand \@@endlink[0]{}%
\providecommand \url  [0]{\begingroup\@sanitize@url \@url }%
\providecommand \@url [1]{\endgroup\@href {#1}{\urlprefix }}%
\providecommand \urlprefix  [0]{URL }%
\providecommand \Eprint [0]{\href }%
\providecommand \doibase [0]{https://doi.org/}%
\providecommand \selectlanguage [0]{\@gobble}%
\providecommand \bibinfo  [0]{\@secondoftwo}%
\providecommand \bibfield  [0]{\@secondoftwo}%
\providecommand \translation [1]{[#1]}%
\providecommand \BibitemOpen [0]{}%
\providecommand \bibitemStop [0]{}%
\providecommand \bibitemNoStop [0]{.\EOS\space}%
\providecommand \EOS [0]{\spacefactor3000\relax}%
\providecommand \BibitemShut  [1]{\csname bibitem#1\endcsname}%
\let\auto@bib@innerbib\@empty
\bibitem [{\citenamefont {v.~Klitzing}\ \emph {et~al.}(1980)\citenamefont
  {v.~Klitzing}, \citenamefont {Dorda},\ and\ \citenamefont
  {Pepper}}]{KlitzingNew1980}%
  \BibitemOpen
  \bibfield  {author} {\bibinfo {author} {\bibfnamefont {K.}~\bibnamefont
  {v.~Klitzing}}, \bibinfo {author} {\bibfnamefont {G.}~\bibnamefont {Dorda}},\
  and\ \bibinfo {author} {\bibfnamefont {M.}~\bibnamefont {Pepper}},\
  }\bibfield  {title} {\bibinfo {title} {New method for high-accuracy
  determination of the fine-structure constant based on quantized hall
  resistance},\ }\href {https://doi.org/10.1103/physrevlett.45.494} {\bibfield
  {journal} {\bibinfo  {journal} {Physical Review Letters}\ }\textbf {\bibinfo
  {volume} {45}},\ \bibinfo {pages} {494} (\bibinfo {year} {1980})}\BibitemShut
  {NoStop}%
\bibitem [{\citenamefont {Tsui}\ \emph {et~al.}(1982)\citenamefont {Tsui},
  \citenamefont {Stormer},\ and\ \citenamefont {Gossard}}]{TsuiTwo1982}%
  \BibitemOpen
  \bibfield  {author} {\bibinfo {author} {\bibfnamefont {D.~C.}\ \bibnamefont
  {Tsui}}, \bibinfo {author} {\bibfnamefont {H.~L.}\ \bibnamefont {Stormer}},\
  and\ \bibinfo {author} {\bibfnamefont {A.~C.}\ \bibnamefont {Gossard}},\
  }\bibfield  {title} {\bibinfo {title} {Two-dimensional magnetotransport in
  the extreme quantum limit},\ }\href
  {https://doi.org/10.1103/physrevlett.48.1559} {\bibfield  {journal} {\bibinfo
   {journal} {Physical Review Letters}\ }\textbf {\bibinfo {volume} {48}},\
  \bibinfo {pages} {1559} (\bibinfo {year} {1982})}\BibitemShut {NoStop}%
\bibitem [{\citenamefont {Hasan}\ and\ \citenamefont
  {Kane}(2010)}]{HasanColloquium2010a}%
  \BibitemOpen
  \bibfield  {author} {\bibinfo {author} {\bibfnamefont {M.~Z.}\ \bibnamefont
  {Hasan}}\ and\ \bibinfo {author} {\bibfnamefont {C.~L.}\ \bibnamefont
  {Kane}},\ }\bibfield  {title} {\bibinfo {title} {Colloquium: {{Topological}}
  insulators},\ }\href {https://doi.org/10.1103/RevModPhys.82.3045} {\bibfield
  {journal} {\bibinfo  {journal} {Reviews of Modern Physics}\ }\textbf
  {\bibinfo {volume} {82}},\ \bibinfo {pages} {3045} (\bibinfo {year}
  {2010})}\BibitemShut {NoStop}%
\bibitem [{\citenamefont {Wen}(2017)}]{WenColloquium2017}%
  \BibitemOpen
  \bibfield  {author} {\bibinfo {author} {\bibfnamefont {X.-G.}\ \bibnamefont
  {Wen}},\ }\bibfield  {title} {\bibinfo {title} {Colloquium: {{Zoo}} of
  quantum-topological phases of matter},\ }\href
  {https://doi.org/10.1103/RevModPhys.89.041004} {\bibfield  {journal}
  {\bibinfo  {journal} {Reviews of Modern Physics}\ }\textbf {\bibinfo {volume}
  {89}},\ \bibinfo {pages} {041004} (\bibinfo {year} {2017})}\BibitemShut
  {NoStop}%
\bibitem [{\citenamefont {Thouless}\ \emph {et~al.}(1982)\citenamefont
  {Thouless}, \citenamefont {Kohmoto}, \citenamefont {Nightingale},\ and\
  \citenamefont {Den~Nijs}}]{ThoulessQuantized1982a}%
  \BibitemOpen
  \bibfield  {author} {\bibinfo {author} {\bibfnamefont {D.~J.}\ \bibnamefont
  {Thouless}}, \bibinfo {author} {\bibfnamefont {M.}~\bibnamefont {Kohmoto}},
  \bibinfo {author} {\bibfnamefont {M.}~\bibnamefont {Nightingale}},\ and\
  \bibinfo {author} {\bibfnamefont {M.}~\bibnamefont {Den~Nijs}},\ }\bibfield
  {title} {\bibinfo {title} {Quantized {{Hall Conductance}} in a
  {{Two-Dimensional Periodic Potential}}},\ }\href
  {https://doi.org/10.1103/PhysRevLett.49.405} {\bibfield  {journal} {\bibinfo
  {journal} {Phys. Rev. Lett.}\ }\textbf {\bibinfo {volume} {49}},\ \bibinfo
  {pages} {405} (\bibinfo {year} {1982})}\BibitemShut {NoStop}%
\bibitem [{\citenamefont {Niu}\ \emph {et~al.}(1985)\citenamefont {Niu},
  \citenamefont {Thouless},\ and\ \citenamefont {Wu}}]{NiuQuantized1985a}%
  \BibitemOpen
  \bibfield  {author} {\bibinfo {author} {\bibfnamefont {Q.}~\bibnamefont
  {Niu}}, \bibinfo {author} {\bibfnamefont {D.~J.}\ \bibnamefont {Thouless}},\
  and\ \bibinfo {author} {\bibfnamefont {Y.-S.}\ \bibnamefont {Wu}},\
  }\bibfield  {title} {\bibinfo {title} {Quantized {{Hall}} conductance as a
  topological invariant},\ }\href {https://doi.org/10.1103/PhysRevB.31.3372}
  {\bibfield  {journal} {\bibinfo  {journal} {Physical Review B}\ }\textbf
  {\bibinfo {volume} {31}},\ \bibinfo {pages} {3372} (\bibinfo {year}
  {1985})}\BibitemShut {NoStop}%
\bibitem [{\citenamefont {Avron}\ and\ \citenamefont
  {Seiler}(1985)}]{AvronQuantization1985}%
  \BibitemOpen
  \bibfield  {author} {\bibinfo {author} {\bibfnamefont {J.~E.}\ \bibnamefont
  {Avron}}\ and\ \bibinfo {author} {\bibfnamefont {R.}~\bibnamefont {Seiler}},\
  }\bibfield  {title} {\bibinfo {title} {Quantization of the hall conductance
  for general, multiparticle schrödinger hamiltonians},\ }\href
  {https://doi.org/10.1103/physrevlett.54.259} {\bibfield  {journal} {\bibinfo
  {journal} {Physical Review Letters}\ }\textbf {\bibinfo {volume} {54}},\
  \bibinfo {pages} {259} (\bibinfo {year} {1985})}\BibitemShut {NoStop}%
\bibitem [{\citenamefont {Xie}\ \emph {et~al.}(2021)\citenamefont {Xie},
  \citenamefont {Pierce}, \citenamefont {Park}, \citenamefont {Parker},
  \citenamefont {Khalaf}, \citenamefont {Ledwith}, \citenamefont {Cao},
  \citenamefont {Lee}, \citenamefont {Chen}, \citenamefont {Forrester},
  \citenamefont {Watanabe}, \citenamefont {Taniguchi}, \citenamefont
  {Vishwanath}, \citenamefont {{Jarillo-Herrero}},\ and\ \citenamefont
  {Yacoby}}]{XieFractional2021}%
  \BibitemOpen
  \bibfield  {author} {\bibinfo {author} {\bibfnamefont {Y.}~\bibnamefont
  {Xie}}, \bibinfo {author} {\bibfnamefont {A.~T.}\ \bibnamefont {Pierce}},
  \bibinfo {author} {\bibfnamefont {J.~M.}\ \bibnamefont {Park}}, \bibinfo
  {author} {\bibfnamefont {D.~E.}\ \bibnamefont {Parker}}, \bibinfo {author}
  {\bibfnamefont {E.}~\bibnamefont {Khalaf}}, \bibinfo {author} {\bibfnamefont
  {P.}~\bibnamefont {Ledwith}}, \bibinfo {author} {\bibfnamefont
  {Y.}~\bibnamefont {Cao}}, \bibinfo {author} {\bibfnamefont {S.~H.}\
  \bibnamefont {Lee}}, \bibinfo {author} {\bibfnamefont {S.}~\bibnamefont
  {Chen}}, \bibinfo {author} {\bibfnamefont {P.~R.}\ \bibnamefont {Forrester}},
  \bibinfo {author} {\bibfnamefont {K.}~\bibnamefont {Watanabe}}, \bibinfo
  {author} {\bibfnamefont {T.}~\bibnamefont {Taniguchi}}, \bibinfo {author}
  {\bibfnamefont {A.}~\bibnamefont {Vishwanath}}, \bibinfo {author}
  {\bibfnamefont {P.}~\bibnamefont {{Jarillo-Herrero}}},\ and\ \bibinfo
  {author} {\bibfnamefont {A.}~\bibnamefont {Yacoby}},\ }\bibfield  {title}
  {\bibinfo {title} {Fractional {{Chern}} insulators in magic-angle twisted
  bilayer graphene},\ }\href {https://doi.org/10.1038/s41586-021-04002-3}
  {\bibfield  {journal} {\bibinfo  {journal} {Nature}\ }\textbf {\bibinfo
  {volume} {600}},\ \bibinfo {pages} {439} (\bibinfo {year}
  {2021})}\BibitemShut {NoStop}%
\bibitem [{\citenamefont {Parameswaran}\ \emph {et~al.}(2013)\citenamefont
  {Parameswaran}, \citenamefont {Roy},\ and\ \citenamefont
  {Sondhi}}]{ParameswaranFractional2013}%
  \BibitemOpen
  \bibfield  {author} {\bibinfo {author} {\bibfnamefont {S.~A.}\ \bibnamefont
  {Parameswaran}}, \bibinfo {author} {\bibfnamefont {R.}~\bibnamefont {Roy}},\
  and\ \bibinfo {author} {\bibfnamefont {S.~L.}\ \bibnamefont {Sondhi}},\
  }\bibfield  {title} {\bibinfo {title} {Fractional quantum {{Hall}} physics in
  topological flat bands},\ }\href {https://doi.org/10.1016/j.crhy.2013.04.003}
  {\bibfield  {journal} {\bibinfo  {journal} {Comptes Rendus Physique}\
  }\textbf {\bibinfo {volume} {14}},\ \bibinfo {pages} {816} (\bibinfo {year}
  {2013})}\BibitemShut {NoStop}%
\bibitem [{\citenamefont {Bergholtz}\ and\ \citenamefont
  {Liu}(2013)}]{BergholtzTopological2013a}%
  \BibitemOpen
  \bibfield  {author} {\bibinfo {author} {\bibfnamefont {E.~J.}\ \bibnamefont
  {Bergholtz}}\ and\ \bibinfo {author} {\bibfnamefont {Z.}~\bibnamefont
  {Liu}},\ }\bibfield  {title} {\bibinfo {title} {Topological flat band models
  and fractional {{Chern}} insulators},\ }\href
  {https://doi.org/10.1142/S021797921330017X} {\bibfield  {journal} {\bibinfo
  {journal} {International Journal of Modern Physics B}\ }\textbf {\bibinfo
  {volume} {27}},\ \bibinfo {pages} {1330017} (\bibinfo {year}
  {2013})}\BibitemShut {NoStop}%
\bibitem [{\citenamefont {Liu}\ and\ \citenamefont
  {Bergholtz}(2023)}]{LiuRecent2023a}%
  \BibitemOpen
  \bibfield  {author} {\bibinfo {author} {\bibfnamefont {Z.}~\bibnamefont
  {Liu}}\ and\ \bibinfo {author} {\bibfnamefont {E.~J.}\ \bibnamefont
  {Bergholtz}},\ }\bibfield  {title} {\bibinfo {title} {Recent developments in
  fractional chern insulators},\ }in\ \href
  {https://doi.org/10.1016/B978-0-323-90800-9.00136-0} {\emph {\bibinfo
  {booktitle} {Reference Module in Materials Science and Materials
  Engineering}}}\ (\bibinfo  {publisher} {Elsevier},\ \bibinfo {year}
  {2023})\BibitemShut {NoStop}%
\bibitem [{\citenamefont {Leykam}\ \emph {et~al.}(2018)\citenamefont {Leykam},
  \citenamefont {Andreanov},\ and\ \citenamefont
  {Flach}}]{LeykamArtificial2018}%
  \BibitemOpen
  \bibfield  {author} {\bibinfo {author} {\bibfnamefont {D.}~\bibnamefont
  {Leykam}}, \bibinfo {author} {\bibfnamefont {A.}~\bibnamefont {Andreanov}},\
  and\ \bibinfo {author} {\bibfnamefont {S.}~\bibnamefont {Flach}},\ }\bibfield
   {title} {\bibinfo {title} {Artificial flat band systems: from lattice models
  to experiments},\ }\href {https://doi.org/10.1080/23746149.2018.1473052}
  {\bibfield  {journal} {\bibinfo  {journal} {Advances in Physics: X}\ }\textbf
  {\bibinfo {volume} {3}},\ \bibinfo {pages} {1473052} (\bibinfo {year}
  {2018})}\BibitemShut {NoStop}%
\bibitem [{\citenamefont {Cao}\ \emph {et~al.}(2018)\citenamefont {Cao},
  \citenamefont {Fatemi}, \citenamefont {Fang}, \citenamefont {Watanabe},
  \citenamefont {Taniguchi}, \citenamefont {Kaxiras},\ and\ \citenamefont
  {{Jarillo-Herrero}}}]{CaoUnconventional2018}%
  \BibitemOpen
  \bibfield  {author} {\bibinfo {author} {\bibfnamefont {Y.}~\bibnamefont
  {Cao}}, \bibinfo {author} {\bibfnamefont {V.}~\bibnamefont {Fatemi}},
  \bibinfo {author} {\bibfnamefont {S.}~\bibnamefont {Fang}}, \bibinfo {author}
  {\bibfnamefont {K.}~\bibnamefont {Watanabe}}, \bibinfo {author}
  {\bibfnamefont {T.}~\bibnamefont {Taniguchi}}, \bibinfo {author}
  {\bibfnamefont {E.}~\bibnamefont {Kaxiras}},\ and\ \bibinfo {author}
  {\bibfnamefont {P.}~\bibnamefont {{Jarillo-Herrero}}},\ }\bibfield  {title}
  {\bibinfo {title} {Unconventional superconductivity in magic-angle graphene
  superlattices},\ }\href {https://doi.org/10.1038/nature26160} {\bibfield
  {journal} {\bibinfo  {journal} {Nature}\ }\textbf {\bibinfo {volume} {556}},\
  \bibinfo {pages} {43} (\bibinfo {year} {2018})}\BibitemShut {NoStop}%
\bibitem [{\citenamefont {Volovik}(2018)}]{VolovikGraphite2018}%
  \BibitemOpen
  \bibfield  {author} {\bibinfo {author} {\bibfnamefont {G.~E.}\ \bibnamefont
  {Volovik}},\ }\bibfield  {title} {\bibinfo {title} {Graphite, {{Graphene}},
  and the {{Flat Band Superconductivity}}},\ }\href
  {https://doi.org/10.1134/S0021364018080052} {\bibfield  {journal} {\bibinfo
  {journal} {JETP Letters}\ }\textbf {\bibinfo {volume} {107}},\ \bibinfo
  {pages} {516} (\bibinfo {year} {2018})}\BibitemShut {NoStop}%
\bibitem [{\citenamefont {Lu}\ \emph {et~al.}(2019)\citenamefont {Lu},
  \citenamefont {Stepanov}, \citenamefont {Yang}, \citenamefont {Xie},
  \citenamefont {Aamir}, \citenamefont {Das}, \citenamefont {Urgell},
  \citenamefont {Watanabe}, \citenamefont {Taniguchi}, \citenamefont {Zhang},
  \citenamefont {Bachtold}, \citenamefont {MacDonald},\ and\ \citenamefont
  {Efetov}}]{LuSuperconductors2019}%
  \BibitemOpen
  \bibfield  {author} {\bibinfo {author} {\bibfnamefont {X.}~\bibnamefont
  {Lu}}, \bibinfo {author} {\bibfnamefont {P.}~\bibnamefont {Stepanov}},
  \bibinfo {author} {\bibfnamefont {W.}~\bibnamefont {Yang}}, \bibinfo {author}
  {\bibfnamefont {M.}~\bibnamefont {Xie}}, \bibinfo {author} {\bibfnamefont
  {M.~A.}\ \bibnamefont {Aamir}}, \bibinfo {author} {\bibfnamefont
  {I.}~\bibnamefont {Das}}, \bibinfo {author} {\bibfnamefont {C.}~\bibnamefont
  {Urgell}}, \bibinfo {author} {\bibfnamefont {K.}~\bibnamefont {Watanabe}},
  \bibinfo {author} {\bibfnamefont {T.}~\bibnamefont {Taniguchi}}, \bibinfo
  {author} {\bibfnamefont {G.}~\bibnamefont {Zhang}}, \bibinfo {author}
  {\bibfnamefont {A.}~\bibnamefont {Bachtold}}, \bibinfo {author}
  {\bibfnamefont {A.~H.}\ \bibnamefont {MacDonald}},\ and\ \bibinfo {author}
  {\bibfnamefont {D.~K.}\ \bibnamefont {Efetov}},\ }\bibfield  {title}
  {\bibinfo {title} {Superconductors, orbital magnets and correlated states in
  magic-angle bilayer graphene},\ }\href
  {https://doi.org/10.1038/s41586-019-1695-0} {\bibfield  {journal} {\bibinfo
  {journal} {Nature}\ }\textbf {\bibinfo {volume} {574}},\ \bibinfo {pages}
  {653} (\bibinfo {year} {2019})}\BibitemShut {NoStop}%
\bibitem [{\citenamefont {Yankowitz}\ \emph {et~al.}(2019)\citenamefont
  {Yankowitz}, \citenamefont {Chen}, \citenamefont {Polshyn}, \citenamefont
  {Zhang}, \citenamefont {Watanabe}, \citenamefont {Taniguchi}, \citenamefont
  {Graf}, \citenamefont {Young},\ and\ \citenamefont
  {Dean}}]{YankowitzTuning2019}%
  \BibitemOpen
  \bibfield  {author} {\bibinfo {author} {\bibfnamefont {M.}~\bibnamefont
  {Yankowitz}}, \bibinfo {author} {\bibfnamefont {S.}~\bibnamefont {Chen}},
  \bibinfo {author} {\bibfnamefont {H.}~\bibnamefont {Polshyn}}, \bibinfo
  {author} {\bibfnamefont {Y.}~\bibnamefont {Zhang}}, \bibinfo {author}
  {\bibfnamefont {K.}~\bibnamefont {Watanabe}}, \bibinfo {author}
  {\bibfnamefont {T.}~\bibnamefont {Taniguchi}}, \bibinfo {author}
  {\bibfnamefont {D.}~\bibnamefont {Graf}}, \bibinfo {author} {\bibfnamefont
  {A.~F.}\ \bibnamefont {Young}},\ and\ \bibinfo {author} {\bibfnamefont
  {C.~R.}\ \bibnamefont {Dean}},\ }\bibfield  {title} {\bibinfo {title} {Tuning
  superconductivity in twisted bilayer graphene},\ }\href
  {https://doi.org/10.1126/science.aav1910} {\bibfield  {journal} {\bibinfo
  {journal} {Science}\ }\textbf {\bibinfo {volume} {363}},\ \bibinfo {pages}
  {1059} (\bibinfo {year} {2019})}\BibitemShut {NoStop}%
\bibitem [{\citenamefont {Peotta}\ and\ \citenamefont
  {T{\"o}rm{\"a}}(2015)}]{PeottaSuperfluidity2015}%
  \BibitemOpen
  \bibfield  {author} {\bibinfo {author} {\bibfnamefont {S.}~\bibnamefont
  {Peotta}}\ and\ \bibinfo {author} {\bibfnamefont {P.}~\bibnamefont
  {T{\"o}rm{\"a}}},\ }\bibfield  {title} {\bibinfo {title} {Superfluidity in
  topologically nontrivial flat bands},\ }\href
  {https://doi.org/10.1038/ncomms9944} {\bibfield  {journal} {\bibinfo
  {journal} {Nature Communications}\ }\textbf {\bibinfo {volume} {6}},\
  \bibinfo {pages} {8944} (\bibinfo {year} {2015})}\BibitemShut {NoStop}%
\bibitem [{\citenamefont {Xie}\ \emph {et~al.}(2020)\citenamefont {Xie},
  \citenamefont {Song}, \citenamefont {Lian},\ and\ \citenamefont
  {Bernevig}}]{XieTopologyBounded2020}%
  \BibitemOpen
  \bibfield  {author} {\bibinfo {author} {\bibfnamefont {F.}~\bibnamefont
  {Xie}}, \bibinfo {author} {\bibfnamefont {Z.}~\bibnamefont {Song}}, \bibinfo
  {author} {\bibfnamefont {B.}~\bibnamefont {Lian}},\ and\ \bibinfo {author}
  {\bibfnamefont {B.~A.}\ \bibnamefont {Bernevig}},\ }\bibfield  {title}
  {\bibinfo {title} {Topology-{{Bounded Superfluid Weight}} in {{Twisted
  Bilayer Graphene}}},\ }\href {https://doi.org/10.1103/PhysRevLett.124.167002}
  {\bibfield  {journal} {\bibinfo  {journal} {Physical Review Letters}\
  }\textbf {\bibinfo {volume} {124}},\ \bibinfo {pages} {167002} (\bibinfo
  {year} {2020})}\BibitemShut {NoStop}%
\bibitem [{\citenamefont {Tasaki}(1998)}]{TasakiNagaoka1998}%
  \BibitemOpen
  \bibfield  {author} {\bibinfo {author} {\bibfnamefont {H.}~\bibnamefont
  {Tasaki}},\ }\bibfield  {title} {\bibinfo {title} {From {{Nagaoka}}'s
  {{Ferromagnetism}} to {{Flat-Band Ferromagnetism}} and {{Beyond}}: {{An
  Introduction}} to {{Ferromagnetism}} in the {{Hubbard Model}}},\ }\href
  {https://doi.org/10.1143/PTP.99.489} {\bibfield  {journal} {\bibinfo
  {journal} {Progress of Theoretical Physics}\ }\textbf {\bibinfo {volume}
  {99}},\ \bibinfo {pages} {489} (\bibinfo {year} {1998})}\BibitemShut
  {NoStop}%
\bibitem [{\citenamefont {Bezrukavnikov}\ and\ \citenamefont
  {Kapustin}(2019)}]{BezrukavnikovLocalization2019}%
  \BibitemOpen
  \bibfield  {author} {\bibinfo {author} {\bibfnamefont {R.}~\bibnamefont
  {Bezrukavnikov}}\ and\ \bibinfo {author} {\bibfnamefont {A.}~\bibnamefont
  {Kapustin}},\ }\bibfield  {title} {\bibinfo {title} {Localization
  {{Properties}} of {{Chern Insulators}}},\ }\href
  {https://doi.org/10.1007/s40598-019-00098-8} {\bibfield  {journal} {\bibinfo
  {journal} {Arnold Mathematical Journal}\ }\textbf {\bibinfo {volume} {5}},\
  \bibinfo {pages} {15} (\bibinfo {year} {2019})}\BibitemShut {NoStop}%
\bibitem [{\citenamefont {Chen}\ \emph {et~al.}(2014)\citenamefont {Chen},
  \citenamefont {Mazaheri}, \citenamefont {Seidel},\ and\ \citenamefont
  {Tang}}]{Chenimpossibility2014}%
  \BibitemOpen
  \bibfield  {author} {\bibinfo {author} {\bibfnamefont {L.}~\bibnamefont
  {Chen}}, \bibinfo {author} {\bibfnamefont {T.}~\bibnamefont {Mazaheri}},
  \bibinfo {author} {\bibfnamefont {A.}~\bibnamefont {Seidel}},\ and\ \bibinfo
  {author} {\bibfnamefont {X.}~\bibnamefont {Tang}},\ }\bibfield  {title}
  {\bibinfo {title} {The impossibility of exactly flat non-trivial {{Chern}}
  bands in strictly local periodic tight binding models},\ }\href
  {https://doi.org/10.1088/1751-8113/47/15/152001} {\bibfield  {journal}
  {\bibinfo  {journal} {Journal of Physics A: Mathematical and Theoretical}\
  }\textbf {\bibinfo {volume} {47}},\ \bibinfo {pages} {152001} (\bibinfo
  {year} {2014})}\BibitemShut {NoStop}%
\bibitem [{\citenamefont {Read}(2017)}]{ReadCompactly2017}%
  \BibitemOpen
  \bibfield  {author} {\bibinfo {author} {\bibfnamefont {N.}~\bibnamefont
  {Read}},\ }\bibfield  {title} {\bibinfo {title} {Compactly supported
  {{Wannier}} functions and algebraic \${{K}}\$-theory},\ }\href
  {https://doi.org/10.1103/PhysRevB.95.115309} {\bibfield  {journal} {\bibinfo
  {journal} {Phys. Rev. B}\ }\textbf {\bibinfo {volume} {95}},\ \bibinfo
  {pages} {115309} (\bibinfo {year} {2017})}\BibitemShut {NoStop}%
\bibitem [{\citenamefont {Kapustin}\ and\ \citenamefont
  {Fidkowski}(2020)}]{KapustinLocal2020}%
  \BibitemOpen
  \bibfield  {author} {\bibinfo {author} {\bibfnamefont {A.}~\bibnamefont
  {Kapustin}}\ and\ \bibinfo {author} {\bibfnamefont {L.}~\bibnamefont
  {Fidkowski}},\ }\bibfield  {title} {\bibinfo {title} {Local {{Commuting
  Projector Hamiltonians}} and the {{Quantum Hall Effect}}},\ }\href
  {https://doi.org/10.1007/s00220-019-03444-1} {\bibfield  {journal} {\bibinfo
  {journal} {Communications in Mathematical Physics}\ }\textbf {\bibinfo
  {volume} {373}},\ \bibinfo {pages} {763} (\bibinfo {year}
  {2020})}\BibitemShut {NoStop}%
\bibitem [{\citenamefont {Sathe}\ \emph {et~al.}(2021)\citenamefont {Sathe},
  \citenamefont {Harper},\ and\ \citenamefont {Roy}}]{SatheCompactly2021}%
  \BibitemOpen
  \bibfield  {author} {\bibinfo {author} {\bibfnamefont {P.}~\bibnamefont
  {Sathe}}, \bibinfo {author} {\bibfnamefont {F.}~\bibnamefont {Harper}},\ and\
  \bibinfo {author} {\bibfnamefont {R.}~\bibnamefont {Roy}},\ }\bibfield
  {title} {\bibinfo {title} {Compactly supported {{Wannier}} functions and
  strictly local projectors},\ }\href
  {https://doi.org/10.1088/1751-8121/ac1167} {\bibfield  {journal} {\bibinfo
  {journal} {Journal of Physics A: Mathematical and Theoretical}\ }\textbf
  {\bibinfo {volume} {54}},\ \bibinfo {pages} {335302} (\bibinfo {year}
  {2021})}\BibitemShut {NoStop}%
\bibitem [{\citenamefont {Sathe}\ and\ \citenamefont
  {Roy}(2023)}]{SatheCompact2023}%
  \BibitemOpen
  \bibfield  {author} {\bibinfo {author} {\bibfnamefont {P.}~\bibnamefont
  {Sathe}}\ and\ \bibinfo {author} {\bibfnamefont {R.}~\bibnamefont {Roy}},\
  }\href {https://doi.org/10.48550/arXiv.2302.11608} {\bibinfo {title} {Compact
  {{Wannier Functions}} in {{One Dimension}}}} (\bibinfo {year} {2023}),\
  \Eprint {https://arxiv.org/abs/2302.11608} {arxiv:2302.11608 [cond-mat,
  physics:math-ph, physics:quant-ph]} \BibitemShut {NoStop}%
\bibitem [{\citenamefont {Danieli}\ \emph {et~al.}(2020)\citenamefont
  {Danieli}, \citenamefont {Andreanov},\ and\ \citenamefont
  {Flach}}]{DanieliManybody2020}%
  \BibitemOpen
  \bibfield  {author} {\bibinfo {author} {\bibfnamefont {C.}~\bibnamefont
  {Danieli}}, \bibinfo {author} {\bibfnamefont {A.}~\bibnamefont {Andreanov}},\
  and\ \bibinfo {author} {\bibfnamefont {S.}~\bibnamefont {Flach}},\ }\bibfield
   {title} {\bibinfo {title} {Many-body flatband localization},\ }\href
  {https://doi.org/10.1103/PhysRevB.102.041116} {\bibfield  {journal} {\bibinfo
   {journal} {Physical Review B}\ }\textbf {\bibinfo {volume} {102}},\ \bibinfo
  {pages} {041116} (\bibinfo {year} {2020})}\BibitemShut {NoStop}%
\bibitem [{\citenamefont {Kuno}\ \emph {et~al.}(2020)\citenamefont {Kuno},
  \citenamefont {Orito},\ and\ \citenamefont {Ichinose}}]{KunoFlatband2020}%
  \BibitemOpen
  \bibfield  {author} {\bibinfo {author} {\bibfnamefont {Y.}~\bibnamefont
  {Kuno}}, \bibinfo {author} {\bibfnamefont {T.}~\bibnamefont {Orito}},\ and\
  \bibinfo {author} {\bibfnamefont {I.}~\bibnamefont {Ichinose}},\ }\bibfield
  {title} {\bibinfo {title} {Flat-band many-body localization and ergodicity
  breaking in the {{Creutz}} ladder},\ }\href
  {https://doi.org/10.1088/1367-2630/ab6352} {\bibfield  {journal} {\bibinfo
  {journal} {New Journal of Physics}\ }\textbf {\bibinfo {volume} {22}},\
  \bibinfo {pages} {013032} (\bibinfo {year} {2020})}\BibitemShut {NoStop}%
\bibitem [{\citenamefont {Danieli}\ \emph {et~al.}(2021)\citenamefont
  {Danieli}, \citenamefont {Andreanov}, \citenamefont {Mithun},\ and\
  \citenamefont {Flach}}]{DanieliNonlinear2021}%
  \BibitemOpen
  \bibfield  {author} {\bibinfo {author} {\bibfnamefont {C.}~\bibnamefont
  {Danieli}}, \bibinfo {author} {\bibfnamefont {A.}~\bibnamefont {Andreanov}},
  \bibinfo {author} {\bibfnamefont {T.}~\bibnamefont {Mithun}},\ and\ \bibinfo
  {author} {\bibfnamefont {S.}~\bibnamefont {Flach}},\ }\bibfield  {title}
  {\bibinfo {title} {Nonlinear caging in all-bands-flat lattices},\ }\href
  {https://doi.org/10.1103/PhysRevB.104.085131} {\bibfield  {journal} {\bibinfo
   {journal} {Physical Review B}\ }\textbf {\bibinfo {volume} {104}},\ \bibinfo
  {pages} {085131} (\bibinfo {year} {2021})}\BibitemShut {NoStop}%
\bibitem [{\citenamefont {Gligori{\'c}}\ \emph {et~al.}(2019)\citenamefont
  {Gligori{\'c}}, \citenamefont {Beli{\v c}ev}, \citenamefont {Leykam},\ and\
  \citenamefont {Maluckov}}]{GligoricNonlinear2019}%
  \BibitemOpen
  \bibfield  {author} {\bibinfo {author} {\bibfnamefont {G.}~\bibnamefont
  {Gligori{\'c}}}, \bibinfo {author} {\bibfnamefont {P.~P.}\ \bibnamefont
  {Beli{\v c}ev}}, \bibinfo {author} {\bibfnamefont {D.}~\bibnamefont
  {Leykam}},\ and\ \bibinfo {author} {\bibfnamefont {A.}~\bibnamefont
  {Maluckov}},\ }\bibfield  {title} {\bibinfo {title} {Nonlinear symmetry
  breaking of {{Aharonov-Bohm}} cages},\ }\href
  {https://doi.org/10.1103/PhysRevA.99.013826} {\bibfield  {journal} {\bibinfo
  {journal} {Physical Review A}\ }\textbf {\bibinfo {volume} {99}},\ \bibinfo
  {pages} {013826} (\bibinfo {year} {2019})}\BibitemShut {NoStop}%
\bibitem [{\citenamefont {Tovmasyan}\ \emph {et~al.}(2018)\citenamefont
  {Tovmasyan}, \citenamefont {Peotta}, \citenamefont {Liang}, \citenamefont
  {T{\"o}rm{\"a}},\ and\ \citenamefont {Huber}}]{TovmasyanPreformed2018}%
  \BibitemOpen
  \bibfield  {author} {\bibinfo {author} {\bibfnamefont {M.}~\bibnamefont
  {Tovmasyan}}, \bibinfo {author} {\bibfnamefont {S.}~\bibnamefont {Peotta}},
  \bibinfo {author} {\bibfnamefont {L.}~\bibnamefont {Liang}}, \bibinfo
  {author} {\bibfnamefont {P.}~\bibnamefont {T{\"o}rm{\"a}}},\ and\ \bibinfo
  {author} {\bibfnamefont {S.~D.}\ \bibnamefont {Huber}},\ }\bibfield  {title}
  {\bibinfo {title} {Preformed pairs in flat {{Bloch}} bands},\ }\href
  {https://doi.org/10.1103/PhysRevB.98.134513} {\bibfield  {journal} {\bibinfo
  {journal} {Physical Review B}\ }\textbf {\bibinfo {volume} {98}},\ \bibinfo
  {pages} {134513} (\bibinfo {year} {2018})}\BibitemShut {NoStop}%
\bibitem [{\citenamefont {Pelegr{\'i}}\ \emph {et~al.}(2020)\citenamefont
  {Pelegr{\'i}}, \citenamefont {Marques}, \citenamefont {Ahufinger},
  \citenamefont {Mompart},\ and\ \citenamefont
  {Dias}}]{PelegriInteractioninduced2020}%
  \BibitemOpen
  \bibfield  {author} {\bibinfo {author} {\bibfnamefont {G.}~\bibnamefont
  {Pelegr{\'i}}}, \bibinfo {author} {\bibfnamefont {A.~M.}\ \bibnamefont
  {Marques}}, \bibinfo {author} {\bibfnamefont {V.}~\bibnamefont {Ahufinger}},
  \bibinfo {author} {\bibfnamefont {J.}~\bibnamefont {Mompart}},\ and\ \bibinfo
  {author} {\bibfnamefont {R.~G.}\ \bibnamefont {Dias}},\ }\bibfield  {title}
  {\bibinfo {title} {Interaction-induced topological properties of two bosons
  in flat-band systems},\ }\href
  {https://doi.org/10.1103/PhysRevResearch.2.033267} {\bibfield  {journal}
  {\bibinfo  {journal} {Physical Review Research}\ }\textbf {\bibinfo {volume}
  {2}},\ \bibinfo {pages} {033267} (\bibinfo {year} {2020})}\BibitemShut
  {NoStop}%
\bibitem [{\citenamefont {Neupert}\ \emph {et~al.}(2011)\citenamefont
  {Neupert}, \citenamefont {Santos}, \citenamefont {Chamon},\ and\
  \citenamefont {Mudry}}]{NeupertFractional2011a}%
  \BibitemOpen
  \bibfield  {author} {\bibinfo {author} {\bibfnamefont {T.}~\bibnamefont
  {Neupert}}, \bibinfo {author} {\bibfnamefont {L.}~\bibnamefont {Santos}},
  \bibinfo {author} {\bibfnamefont {C.}~\bibnamefont {Chamon}},\ and\ \bibinfo
  {author} {\bibfnamefont {C.}~\bibnamefont {Mudry}},\ }\bibfield  {title}
  {\bibinfo {title} {Fractional {{Quantum Hall States}} at {{Zero Magnetic
  Field}}},\ }\href {https://doi.org/10.1103/PhysRevLett.106.236804} {\bibfield
   {journal} {\bibinfo  {journal} {Physical Review Letters}\ }\textbf {\bibinfo
  {volume} {106}},\ \bibinfo {pages} {236804} (\bibinfo {year}
  {2011})}\BibitemShut {NoStop}%
\bibitem [{\citenamefont {Sun}\ \emph {et~al.}(2011)\citenamefont {Sun},
  \citenamefont {Gu}, \citenamefont {Katsura},\ and\ \citenamefont
  {Das~Sarma}}]{SunNearly2011}%
  \BibitemOpen
  \bibfield  {author} {\bibinfo {author} {\bibfnamefont {K.}~\bibnamefont
  {Sun}}, \bibinfo {author} {\bibfnamefont {Z.}~\bibnamefont {Gu}}, \bibinfo
  {author} {\bibfnamefont {H.}~\bibnamefont {Katsura}},\ and\ \bibinfo {author}
  {\bibfnamefont {S.}~\bibnamefont {Das~Sarma}},\ }\bibfield  {title} {\bibinfo
  {title} {Nearly {{Flatbands}} with {{Nontrivial Topology}}},\ }\href
  {https://doi.org/10.1103/PhysRevLett.106.236803} {\bibfield  {journal}
  {\bibinfo  {journal} {Physical Review Letters}\ }\textbf {\bibinfo {volume}
  {106}},\ \bibinfo {pages} {236803} (\bibinfo {year} {2011})}\BibitemShut
  {NoStop}%
\bibitem [{\citenamefont {Tang}\ \emph {et~al.}(2011)\citenamefont {Tang},
  \citenamefont {Mei},\ and\ \citenamefont {Wen}}]{TangHighTemperature2011a}%
  \BibitemOpen
  \bibfield  {author} {\bibinfo {author} {\bibfnamefont {E.}~\bibnamefont
  {Tang}}, \bibinfo {author} {\bibfnamefont {J.-W.}\ \bibnamefont {Mei}},\ and\
  \bibinfo {author} {\bibfnamefont {X.-G.}\ \bibnamefont {Wen}},\ }\bibfield
  {title} {\bibinfo {title} {High-{{Temperature Fractional Quantum Hall
  States}}},\ }\href {https://doi.org/10.1103/PhysRevLett.106.236802}
  {\bibfield  {journal} {\bibinfo  {journal} {Physical Review Letters}\
  }\textbf {\bibinfo {volume} {106}},\ \bibinfo {pages} {236802} (\bibinfo
  {year} {2011})}\BibitemShut {NoStop}%
\bibitem [{\citenamefont {Wang}\ and\ \citenamefont
  {Ran}(2011)}]{WangNearly2011}%
  \BibitemOpen
  \bibfield  {author} {\bibinfo {author} {\bibfnamefont {F.}~\bibnamefont
  {Wang}}\ and\ \bibinfo {author} {\bibfnamefont {Y.}~\bibnamefont {Ran}},\
  }\bibfield  {title} {\bibinfo {title} {Nearly flat band with {{Chern}} number
  \${{C}}=2\$ on the dice lattice},\ }\href
  {https://doi.org/10.1103/PhysRevB.84.241103} {\bibfield  {journal} {\bibinfo
  {journal} {Physical Review B}\ }\textbf {\bibinfo {volume} {84}},\ \bibinfo
  {pages} {241103} (\bibinfo {year} {2011})}\BibitemShut {NoStop}%
\bibitem [{\citenamefont {Scaffidi}\ and\ \citenamefont
  {Simon}(2014)}]{ScaffidiExact2014a}%
  \BibitemOpen
  \bibfield  {author} {\bibinfo {author} {\bibfnamefont {T.}~\bibnamefont
  {Scaffidi}}\ and\ \bibinfo {author} {\bibfnamefont {S.~H.}\ \bibnamefont
  {Simon}},\ }\bibfield  {title} {\bibinfo {title} {Exact solutions of
  fractional {{Chern}} insulators: {{Interacting}} particles in the
  {{Hofstadter}} model at finite size},\ }\href
  {https://doi.org/10.1103/PhysRevB.90.115132} {\bibfield  {journal} {\bibinfo
  {journal} {Physical Review B}\ }\textbf {\bibinfo {volume} {90}},\ \bibinfo
  {pages} {115132} (\bibinfo {year} {2014})}\BibitemShut {NoStop}%
\bibitem [{Note1()}]{Note1}%
  \BibitemOpen
  \bibinfo {note} {Strictly speaking, in addition to the condition $L_{x,y}
  \geq 3nR$, we also require that the Hamiltonian satisfy $L_{x,y} \geq 2nR +2$
  to ensure that all the projectors are SL. Except for the special (and
  uninteresting) case of $nR\in \{0,1\}$, it is always true that $3nR \geq 2nR
  +2$, so that we simply require $L_{x,y} \geq 3nR$.}\BibitemShut {Stop}%
\bibitem [{Note2()}]{Note2}%
  \BibitemOpen
  \bibinfo {note} {In Ref.~\cite {DanieliNonlinear2021}, it was stated without
  proof that 1d SL flat Hamiltonians with LTI must have SL band projectors.
  [See the discussion below equation (A22) in that paper.] While this statement
  is true, we believe that it is not trivial to prove. Hence by proving this
  statement, we fill a gap in their proof for the statement that each band in a
  1d flat Hamiltonian can be spanned by compact Wannier functions. Furthermore,
  our proof regarding the projectors is not restricted to 1d systems, or to
  systems with LTI.}\BibitemShut {Stop}%
\bibitem [{Note3()}]{Note3}%
  \BibitemOpen
  \bibinfo {note} {Adding a constant to the Hamiltonian has no effect on the
  projectors.}\BibitemShut {Stop}%
\bibitem [{Note4()}]{Note4}%
  \BibitemOpen
  \bibinfo {note} {If $L_x$ is not divisible by $q$, then we choose the first
  $\protect \floor *{L_x/q}-1$ number of supercells along $\protect \hat x$ to
  consist of $q$ cells each (where $\protect \floor *{.}$ denotes the floor
  function), while the last supercell will comprise $L_x - q (\protect \floor
  *{L_x/q}-1)$ number of primitive cells. We implement a similar grouping along
  $\protect \hat y$ if $L_y$ is not divisible by $q$.}\BibitemShut {Stop}%
\bibitem [{\citenamefont {Peierls}(1933)}]{PeierlsZur1933}%
  \BibitemOpen
  \bibfield  {author} {\bibinfo {author} {\bibfnamefont {R.}~\bibnamefont
  {Peierls}},\ }\bibfield  {title} {\bibinfo {title} {{Zur Theorie des
  Diamagnetismus von Leitungselektronen}},\ }\href
  {https://doi.org/10.1007/BF01342591} {\bibfield  {journal} {\bibinfo
  {journal} {Zeitschrift f\"ur Physik}\ }\textbf {\bibinfo {volume} {80}},\
  \bibinfo {pages} {763} (\bibinfo {year} {1933})}\BibitemShut {NoStop}%
\bibitem [{\citenamefont {Hastings}\ and\ \citenamefont
  {Michalakis}(2015)}]{HastingsQuantization2015}%
  \BibitemOpen
  \bibfield  {author} {\bibinfo {author} {\bibfnamefont {M.~B.}\ \bibnamefont
  {Hastings}}\ and\ \bibinfo {author} {\bibfnamefont {S.}~\bibnamefont
  {Michalakis}},\ }\bibfield  {title} {\bibinfo {title} {Quantization of {{Hall
  Conductance}} for {{Interacting Electrons}} on a {{Torus}}},\ }\href
  {https://doi.org/10.1007/s00220-014-2167-x} {\bibfield  {journal} {\bibinfo
  {journal} {Communications in Mathematical Physics}\ }\textbf {\bibinfo
  {volume} {334}},\ \bibinfo {pages} {433} (\bibinfo {year}
  {2015})}\BibitemShut {NoStop}%
\bibitem [{\citenamefont {Kudo}\ \emph {et~al.}(2019)\citenamefont {Kudo},
  \citenamefont {Watanabe}, \citenamefont {Kariyado},\ and\ \citenamefont
  {Hatsugai}}]{KudoManyBody2019}%
  \BibitemOpen
  \bibfield  {author} {\bibinfo {author} {\bibfnamefont {K.}~\bibnamefont
  {Kudo}}, \bibinfo {author} {\bibfnamefont {H.}~\bibnamefont {Watanabe}},
  \bibinfo {author} {\bibfnamefont {T.}~\bibnamefont {Kariyado}},\ and\
  \bibinfo {author} {\bibfnamefont {Y.}~\bibnamefont {Hatsugai}},\ }\bibfield
  {title} {\bibinfo {title} {Many-{{Body Chern Number}} without
  {{Integration}}},\ }\href {https://doi.org/10.1103/PhysRevLett.122.146601}
  {\bibfield  {journal} {\bibinfo  {journal} {Physical Review Letters}\
  }\textbf {\bibinfo {volume} {122}},\ \bibinfo {pages} {146601} (\bibinfo
  {year} {2019})}\BibitemShut {NoStop}%
\bibitem [{Note5()}]{Note5}%
  \BibitemOpen
  \bibinfo {note} {Let us note that while our proof of Chern triviality was
  stated for flat bands in flat Hamiltonians, all the steps can also be used to
  prove that the Chern number associated with an isolated degenerate energy (or
  a flat band) that has SL projector is zero, regardless of whether the rest of
  the bands are flat.}\BibitemShut {Stop}%
\end{thebibliography}%

\clearpage
\newpage

\setcounter{equation}{0}
\setcounter{figure}{0}
\setcounter{section}{0}
\setcounter{table}{0}
\setcounter{page}{1}
\makeatletter
\renewcommand{\theequation}{S\arabic{equation}}
\renewcommand{\thepage}{S-\arabic{page}}
\renewcommand{\thesection}{S\arabic{section}}

\begin{widetext}
\section*{Supplementary Material to ``Topological Triviality of Flat Hamiltonians''}

\section{Proof of Spectrum not changing after Flux Insertion}\label{appendix:unchanging_spectrum_proof}
In the main text, it was asserted that when a flux $\phi_x$ is inserted, the spectrum of a flat Hamiltonian does not change (if $L_x \geq 3q$ and $L_x\geq 2q+2$).
Here, we provide a proof for this statement.

Let us denote the spectral decomposition of the flat Hamiltonian $H$ by
\begin{align}
    H = \sum_{i=1}^n E_i P_i,
\end{align}
where $P_i P_j =P_j P_i = P_i \delta_{ij}$.
$R$ and $p_i$ denote the maximum hopping ranges of $H$ and $P_i$ respectively, and $q\coloneqq \max \{R, p_1,\dotsc, p_n \}$.
As described in the main text, the Hamiltonian gets transformed according to Peierls substitution upon inserting $\phi_x$.
We chose a gauge for the substitution which corresponds to twisted boundary conditions (TBCs) with a boundary at $x=L_x$ (or equivalently, $x=0$).
In order to implement Peierls substitution on the projectors as well, we require that they be SL along $\hat x$.
This requires that $L_x \geq 2q+2$.
Denoting by $P_i(\phi_x)$ the post-substitution version of $P_i$, we obtain
\begin{align}
    H(\phi_x) =  \sum_{i=1}^n E_i P_i (\phi_x). \label{eq:H_after_phi_x_appendix}
\end{align}
We will now show that similar to the original $P_i$s, each $P_i(\phi_x)$ is an orthogonal projector and that any two distinct projectors from this transformed set are mutually orthogonal.

We will find it convenient to work in the size $q\times q$ supercell representation described in the main text, so that $H$ and all the $P_i$s are NNN-hopping.
Since the $y$-coordinate does not play a role here, the problem is effectively 1d.
Hence, we suppress the $y$-coordinate for the following steps.
Each $P_i$ can then effectively be treated as a 1d NN projector.

We will start by showing that each $P_i(\phi_x)$ is an orthogonal projector.
To that end, let us consider any projector $P \in \{P_1,\dotsc,P_n\}$.
If $L_x \geq 3q$, then there are at least $3$ supercells in the lattice.
Peierls substitution for $P$ (implemented in the primitive cell representation) is then equivalent to modifying only those hopping matrices [as defined in Eq.~\eqref{eq:defn_hopping_matrices} in the main text] that connect $x=0$ and $x=1$ in the supercell representation.
Thus, in the supercell representation, the changes are captured by the following transformations of the hopping matrices:
\begin{eqn}
        \p 1 0 &\xrightarrow{\phi_x} e^{-2\pi i \phi_x/\Phi_0}\p 1 0 \\
        \text{and }\p {-1} 1 &\xrightarrow{\phi_x}  e^{2\pi i \phi_x/\Phi_0}\p {-1} 1. \label{eq:projector_change_after_phi_appendix}
\end{eqn}
Note that if $L_x <3q$, then we only have two supercells, and Peierls substitution implemented in the primitive cell representation is not equivalent to \eqref{eq:projector_change_after_phi_appendix} in the supercell representation.
Since we make use of \eqref{eq:projector_change_after_phi_appendix} below, we require that $L_x \geq 3q$ at this point.

Before inserting a flux, $P$ is an orthogonal projector and hence satisfies $P^2=P=P^\dagger$. 
This is equivalent to the conditions
\begin{eqn}
        (\p j x)^\dagger &= \p {-j} {x+j}, \\
        \p 0 x &= (\p 0 x)^2 +  (\p 1 x)^\dagger (\p 1 x) + (\p 1 {x-1})   (\p 1 {x-1})^\dagger, \\
        \text{and } \p 1 x &= \p 0 {x+1} \p 1 x + \p 1 x \p 0 x, \label{eq:hopping_conditions}
\end{eqn}
for all cell positions $x$.
It is straightforward to show $P$ continues to satisfy \eqref{eq:hopping_conditions} even after undergoing Peierls substitution \eqref{eq:projector_change_after_phi_appendix}.
Thus, every $P_i(\phi_x)$ is also an orthogonal projector.

Now let us consider any two distinct projectors $P,Q \in \{ P_1,\dotsc,P_n\}$. 
They necessarily satisfy $PQ=QP=0$. 
Expressed in terms of the hopping matrices, these equalities can be written as
\begin{eqn}
    \p 0 x Q_0^x = Q_0^x P_0^x &= 0 \\
    P_1^x Q_0 ^x + P_0^{x+1} Q_1^x = Q_1^x P_0^x + Q_0^{x+1} P_1^x &= 0 \\
    P_1^{x+1} Q_1^x = Q_1^{x+1} P_1^x &=0. \label{eq:mutual_ortho}
\end{eqn}
It is easy to show that $P$ and $Q$ continue to satisfy \eqref{eq:mutual_ortho} even after undergoing the substitution \eqref{eq:projector_change_after_phi_appendix} (and its analog for $Q$).
Thus, all the $P_i(\phi_x)$s together are a set of mutually orthogonal projectors.
Eq.~\eqref{eq:H_after_phi_x_appendix} is therefore a spectral decomposition.

\end{widetext}
\end{document}